\documentclass[prl,superscriptaddress,showpacs,twocolumn,10pt]{revtex4}

\usepackage{amssymb}
\usepackage{amsmath}
\usepackage{mathrsfs}
\usepackage{graphicx}
\usepackage{epic}
\usepackage{eepic}

\newtheorem{theorem}{Theorem}
\newtheorem{lemma}{Lemma}

\newtheorem{corollary}{Corollary}

\begin{document}

\title{Realization of positive-operator-valued measures by projective measurements without introducing ancillary dimensions}
\author{Guoming Wang}
\email{wgm00@mails.tsinghua.edu.cn}
\author{Mingsheng Ying}
\email{yingmsh@tsinghua.edu.cn}

\affiliation{State Key Laboratory of Intelligent Technology and
Systems, Department of Computer Science and Technology, Tsinghua
University, Beijing, China, 100084}

\date{\today}

\begin{abstract}

We propose a scheme that can realize a class of
positive-operator-valued measures (POVMs) by performing a sequence
of projective measurements on the original system, in the sense
that for an arbitrary input state the probability distribution of
the measurement outcomes is faithfully reproduced. A necessary and
sufficient condition for a POVM to be realizable in this way is
also derived. In contrast to the canonical approach provided by
Neumark's theorem, our method has the advantage of requiring no
auxiliary system. Moreover, an arbitrary POVM can be realized by
utilizing our protocol on an extended space which is formed by
adding only a single extra dimension.

\end{abstract}


\maketitle

Realization of generalized quantum measurements, or
positive-operator-valued measures (POVMs), is a fundamental
problem in quantum mechanics and quantum information science. Many
tasks, such as quantum state discrimination and entanglement
transformation, require nonorthogonal measurements to success or
to achieve the optimal efficiency. During recent years much effort
has been devoted to the implementation of POVMs on various kinds
of physical systems \cite{EFFORT}. Many of the proposed schemes
are derived from Neumark's theorem \cite{AP93}, which asserts that
any POVM can be realized by extending the original Hilbert space
to a larger space and performing a projective measurement on the
extended space. In spite of its universality, this method has the
drawback of needing a collective operation on the original system
and an ancillary system, which may be difficult to implement in
practice.

In this letter, we provide a scheme that can realize a class of
POVMs by performing a series of projective measurements on the
original system alone, in the sense that for an arbitrary input
state the probability distribution of the measurement outcomes is
faithfully reproduced. A necessary and sufficient condition for a
POVM to be realizable in this way is also derived. Moreover, if
only a single ancillary dimension is introduced, then arbitrary
POVMs can be realized by applying our protocol on the enlarged
space. Nevertheless, our method is limited to the physical systems
which can be repeatedly measured, i.e. we have access to the
post-measurement states and can perform operations on them again.

Let us begin with the following example. Suppose we are given an
arbitrary three-dimensional state, i.e. a qutrit. Consider the
following protocol:

(1)Perform the projective measurement
$\{|0\rangle\langle0|+|1\rangle\langle1|,|2\rangle\langle2|\}$ on
the initial state. Suppose the outcome $r=0/1$ or $2$ is obtained.

(2)If $r=2$, do nothing. If $r=0/1$, perform the projective
measurement
$\{|\phi_0\rangle\langle\phi_0|,|\phi_1\rangle\langle\phi_1|,|\phi_2\rangle\langle\phi_2|\}$
on the state after stage (1), where
$|\phi_0\rangle=\frac{1}{\sqrt{3}}(|0\rangle+|1\rangle+|2\rangle)$,
$|\phi_1\rangle=\frac{1}{\sqrt{14}}(|0\rangle+2|1\rangle-3|2\rangle)$,
$|\phi_2\rangle=\frac{1}{\sqrt{42}}(5|0\rangle-4|1\rangle-|2\rangle)$.
Suppose the outcome $s=\phi_0$, $\phi_1$ or $\phi_2$ is obtained.

(3) If $r=0/1$ and $s=\phi_0$, then output the final outcome $0$;
if $r=0/1$ and $s=\phi_1$, then output the final outcome $1$;
otherwise, output the final outcome $2$.

Denoting the initial state by $\rho$, one can verify that the
probabilities of obtaining the final outcomes $0$,$1$ and $2$ are
$tr(\rho E_{0}),tr(\rho E_{1}),tr(\rho E_{2})$ respectively, where
$E_0=\frac{2}{3}|\psi_0\rangle\langle \psi_0|$,
$E_1=\frac{5}{14}|\psi_1\rangle\langle \psi_1|$, $E_2=I-E_0-E_1$
with $|\psi_0\rangle=\frac{1}{\sqrt{2}}(|0\rangle+|1\rangle)$,
$|\psi_1\rangle=\frac{1}{\sqrt{5}}(|0\rangle+2|1\rangle)$. Thus,
the above protocol actually can be viewed as a `virtual' POVM
$\{E_0,E_1,E_2\}$ if only the measurement outcome is concerned.

\begin{figure}[ht]
\begin{center}
\setlength{\unitlength}{0.2cm}
\begin{picture}(30,22)
\thicklines

\drawline(15,20)(10,12) \drawline(15,20)(20,12)
\drawline(10,12)(5,4) \drawline(10,12)(10,4)
\drawline(10,12)(15,4)

\put(3,0){\makebox(5,5){0}} \put(8,0){\makebox(5,5){1}}
\put(13,0){\makebox(5,5){2}} \put(18,8){\makebox(5,5){2}}

\put(5,4){\circle*{1}} \put(10,4){\circle*{1}}
\put(15,4){\circle*{1}}
 \put(10,12){\circle*{1}}
\put(20,12){\circle*{1}} \put(15,20){\circle*{1}}

\put(8,14){\makebox(5,5){0/1}} \put(16,14){\makebox(5,5){2}}
\put(3,5){\makebox(5,5){$\phi_0$}}
\put(8.5,5){\makebox(5,5){$\phi_1$}}
\put(12,5){\makebox(5,5){$\phi_2$}}

\end{picture}
\caption{A graphical depiction of the example protocol. The
numbers $0/1,2$ and symbols $\phi_0,\phi_1,\phi_2$ along the
branches represent the possible outcomes of the measurements. The
numbers $0$,$1$ and $2$ beside the leaf nodes indicate the final
outcomes inferred from the corresponding chain of measurements.}
\label{fig1}
\end{center}
\end{figure}
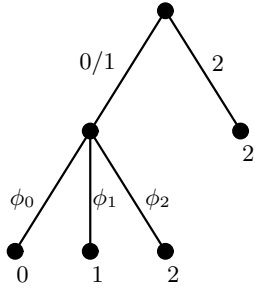

The above example demonstrated our basic ideas. Instead of
introducing an auxiliary system and performing a collective
projective measurement on the extended system, we here repeat
projective measurements on the original system over and over
again. Suppose we first perform a projective measurement
$\{P_1,\dots,P_m\}$ on the initial state and the outcome $i$ is
obtained. Then depending on $i$ we choose another projective
measurement $\{P^{(i)}_1,\dots,P^{(i)}_{m_{(i)}}\}$ and perform it
on the state after the first measurement. Suppose its outcome is
$i_1$. Then basing on $i$ and $i_1$ we construct another
projective measurement
$\{P^{(i,i_1)}_1,\dots,P^{(i,i_1)}_{m_{(i,i_1)}}\}$ and perform it
on the state after the second measurement. Similarly the protocol
goes on. When getting every possible outcome of a performed
measurement, we may either to output a final outcome and finish,
or to proceed with a new measurement. All these should be
specified in advance by the protocol and should not depend on the
input state. If such a protocol outputs the final outcomes with
the same probability distribution as that of a POVM for an
arbitrary input state, we say that this POVM is realized by this
protocol, although in our method the final outcome is created more
artificially. It is worth noting that every protocol has a concise
graphical depiction in the form of a tree structure (For detailed
definitions of graphs and trees, see Ref.\cite{TREE}). Each leaf
node represents a possible outlet of the protocol and has a
corresponding final outcome. For the above example, its protocol
tree is shown in Fig.\ref{fig1}.

In general, we can come up with much more complex protocols than
the one given above. It is easy to observe that any such protocol
will generate a POVM. Conversely, we may ask whether an arbitrary
POVM can be realized in this way. To understand the limit of our
method, suppose that a POVM $\{E_1,\dots,E_m\}$ is realized by a
protocol which begins with a projective measurement
$\{P_1,\dots,P_n\}$. No matter how the protocol works specifically
in the subsequent steps, it is always true that every POVM element
$E_k$ should be written as the sum of some items $M^{\dagger}M$
where
\begin{equation}
M=P^{(i,i_1,\dots,i_{t-1})}_{i_t} \dots P^{(i,i_1)}_{i_2}
P^{(i)}_{i_1} P_i
\end{equation}
is the product of the projection operators in a chain of
measurements. Then we have $M^{\dagger}M=P_i\Pi P_i$ for some
positive operator $\Pi$. Thus, $E_k$ could be written in the form
$E_k=\sum_{i=1}^{n}{P_i \Omega_{ki} P_i}$ for some positive
operators $\Omega_{ki}$. It then follows that
\begin{equation}
E_k=\sum_{i=1}^{n}{P_i E_k P_i}.
\label{eq:eesp}
\end{equation}
As a consequence, for each $E_k$ and each $P_i$, we have
$[E_k,P_i]=0$. So a necessary condition for a POVM to be
realizable by our approach is that there exists at least one
projection operator $P \neq I$ such that $P$ commutes with all the
POVM elements.

Next we will prove that the above condition is also sufficient.
Suppose a POVM $\{E_1,...,E_m\}$ and a projection operator $P \neq
I$ satisfy that $[E_i,P]=0$ for $i=1,\dots,m$. Let $E_{i0}=PE_iP$,
$E_{i1}=(I-P)E_i(I-P)$, and suppose they have spectral
decompositions
\begin{equation}
E_{ia}=\sum_{j=1}^{d_{ia}}{\lambda^{j}_{ia}|\phi^{j}_{ia}\rangle\langle
\phi^{j}_{ia}|}, \label{eq:spec}
\end{equation}
where $a=0,1$, $i=1,\dots,m$ and $d_{ia}=rank(E_{ia})$. One can
verify that $E_i=E_{i0}+E_{i1}$, $\sum_{i=1}^{m}E_{i0}=P$ and
$\sum_{i=1}^{m}E_{i1}=I-P$.

Before presenting our protocol for realizing this POVM, a lemma
should be stated first:
\begin{lemma}
If a linear operator $M$, a state $|\phi\rangle$ and a number
$\lambda >0$ satisfy $M^{\dagger}M-\lambda|\phi\rangle\langle\phi|
\ge 0$, then there exist a state $|\theta\rangle$ and a number
$\mu \ge \lambda$ such that
\begin{equation}
M^{\dagger}|\theta\rangle=\sqrt{\mu}|\phi\rangle. \label{eq:lem}
\end{equation}
\label{lemma1}
\end{lemma}

\textit{Proof.} The proof is given in the appendix. $\square$

The constructive proof of this lemma gives us a basic function
which takes $M$, $\lambda$ and $|\phi\rangle$ as input and outputs
$\mu$, $|\theta\rangle$ in the Eq.(\ref{eq:lem}). We write it in
the form
\begin{equation}
f(M,\lambda, |\phi\rangle)=(\mu,|\theta\rangle).
\end{equation}

Our protocol is as follows:

Stage 1: Perform the projective measurement $\{P, I-P\}$ on the
initial state. If the outcome corresponding to $P$ is obtained,
set $a=0$; otherwise, set $a=1$.

Stage 2:

(2.1) Set $i=1$, $j=1$. If $a=0$, set $M^{1}_{10}=P$; otherwise,
set $M^{1}_{11}=I-P$.

(2.2) Compute the function
\begin{equation}
f(M^{j}_{ia},
\lambda^{j}_{ia},|\phi^{j}_{ia}\rangle)=(\mu^{j}_{ia},|\theta^{j}_{ia}\rangle).
\label{eq:functionf}
\end{equation}

(2.3) Choose a state $|\xi^{j}_{ia}\rangle \in
ker(M^{j\dagger}_{ia})$. (This is always possible, and we will
prove it later.)

(2.4) Perform the projective measurement
$\{|\psi^{j}_{ia}\rangle\langle\psi^{j}_{ia}|,
I-|\psi^{j}_{ia}\rangle\langle\psi^{j}_{ia}|\}$ on the current
state, where
\begin{eqnarray}
|\psi^{j}_{ia}\rangle=\sqrt{\frac{\lambda^{j}_{ia}}{\mu^{j}_{ia}}}|\theta^{j}_{ia}\rangle
+\sqrt{1-\frac{\lambda^{j}_{ia}}{\mu^{j}_{ia}}}|\xi^{j}_{ia}\rangle.
\label{eq:psijia}
\end{eqnarray}
If the outcome corresponding to $|\psi^{j}_{ia}\rangle$ is
obtained, then output the final outcome $i$ and exit; otherwise,
goto stage (2.5).

(2.5)If $j<d_{ia}$, then set
\begin{equation}
M^{j+1}_{ia}=(I-
|\psi^{j}_{ia}\rangle\langle\psi^{j}_{ia}|)M^{j}_{ia},
\label{eq:2.5.1}
\end{equation}
and increase $j$ by 1; otherwise, set
\begin{equation}
M^{1}_{(i+1)a}=(I-
|\psi^{j}_{ia}\rangle\langle\psi^{j}_{ia}|)M^{j}_{ia},
\label{eq:2.5.2}
\end{equation}
increase $i$ by 1, and set $j=1$.

(2.6)If $i=m$, then output the final outcome $m$ and exit;
otherwise goto stage (2.2).

\begin{figure}[ht]
\begin{center}
\setlength{\unitlength}{0.2cm}
\begin{picture}(42,40)
\thicklines

\drawline(21,38)(17,30) \drawline(21,38)(25,30)

\drawline(17,30)(14,24) \drawline(17,30)(20,24)
\drawline(25,30)(22,24) \drawline(25,30)(28,24)

\drawline(14,24)(11,18) \drawline(14,24)(17,18)
\drawline(28,24)(25,18) \drawline(28,24)(31,18)

\dottedline[$\bullet$]{1.5}(11,17)(9,13)
\dottedline[$\bullet$]{1.5}(31,17)(33,13)

\drawline(8,12)(5,6) \drawline(8,12)(11,6) \drawline(34,12)(31,6)
\drawline(34,12)(37,6)

\put(21,38){\circle*{1}} \put(17,30){\circle*{1}}
\put(25,30){\circle*{1}} \put(14,24){\circle*{1}}
\put(20,24){\circle*{1}} \put(28,24){\circle*{1}}
\put(22,24){\circle*{1}} \put(11,18){\circle*{1}}
\put(17,18){\circle*{1}} \put(25,18){\circle*{1}}
\put(31,18){\circle*{1}} \put(8,12){\circle*{1}}
\put(34,12){\circle*{1}} \put(5,6){\circle*{1}}
\put(11,6){\circle*{1}} \put(31,6){\circle*{1}}
\put(37,6){\circle*{1}}

\put(16,33){\makebox(5,5){$P$}} \put(22,33){\makebox(5,5){$I-P$}}
\put(17,26){\makebox(5,5){$\psi^{1}_{10}$}}
\put(20,26){\makebox(5,5){$\psi^{1}_{11}$}}
\put(14,20){\makebox(5,5){$\psi^{2}_{10}$}}
\put(23,20){\makebox(5,5){$\psi^{2}_{11}$}}
\put(11,7){\makebox(5,5){$\psi^{d_{(m-1)0}}_{(m-1)0}$}}
\put(27,7){\makebox(5,5){$\psi^{d_{(m-1)1}}_{(m-1)1}$}}

\put(8,26){\makebox(5,5){$I-|\psi^{1}_{10}\rangle\langle\psi^{1}_{10}|$}}
\put(5,20){\makebox(5,5){$I-|\psi^{2}_{10}\rangle\langle\psi^{2}_{10}|$}}
\put(29,26){\makebox(5,5){$I-|\psi^{1}_{11}\rangle\langle\psi^{1}_{11}|$}}
\put(32,20){\makebox(5,5){$I-|\psi^{2}_{11}\rangle\langle\psi^{2}_{11}|$}}
\put(2,7){\makebox(5,5){$Q_0$}} \put(35,7){\makebox(5,5){$Q_1$}}

\put(17,20){\makebox(5,5){$1$}} \put(20,20){\makebox(5,5){$1$}}
\put(14,14){\makebox(5,5){$1$}} \put(23,14){\makebox(5,5){$1$}}
\put(8,2){\makebox(5,5){$m-1$}} \put(29,2){\makebox(5,5){$m-1$}}
\put(2,2){\makebox(5,5){$m$}} \put(35,2){\makebox(5,5){$m$}}

\end{picture}
\caption{A graphical depiction of the general protocol. The dotted
lines stand for omitted branches. For the measurement outcome
corresponding to the state $|\psi^{j}_{ia}\rangle$ at every
iteration of stage (2.4), the symbol $\psi^{j}_{ia}$ is put along
the branch; and for the other outcome, we put the performed linear
operation $I-|\psi^{j}_{ia}\rangle\langle\psi^{j}_{ia}|$ along the
other branch, including
$Q_a=I-|\psi^{d_{(m-1)a}}_{(m-1)a}\rangle\langle\psi^{d_{(m-1)a}}_{(m-1)a}|$
where $a=0,1$. The numbers $1$,$\dots$,$m-1$ and $m$ beside the
leaf nodes indicate the final outcomes inferred from the
corresponding chain of measurements. }

\label{fig2}
\end{center}
\end{figure}
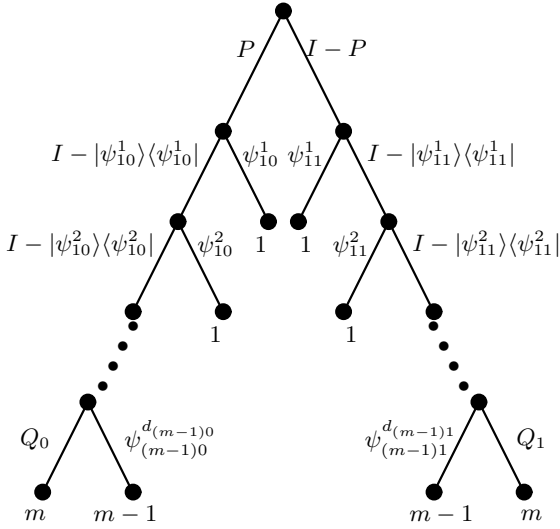

The protocol can be depicted by the tree shown in Fig.\ref{fig2}.
One can see that its structure is really simple.

To prove the validity of the protocol, it suffices to consider
pure input states since the probability of obtaining each
measurement outcome is linear in the density matrix of the input
state. Suppose the input state is $|\psi\rangle$. It can be
observed that the role of $M^{j}_{ia}$ is to record the total
operation performed on the input state, which means, no matter at
what stage of the protocol, the current state is always
$|\psi'\rangle={M^{j}_{ia}|\psi\rangle}/{\|M^{j}_{ia}|\psi\rangle\|}$
and the protocol can reach this stage with probability
$\|M^{j}_{ia}|\psi\rangle\|^{2}$.

Since every $M^{j}_{ia}$ is the product of a chain of projection
operators which start with either $P$ or $I-P$, we have
$rank(M^{j}_{ia}) \le rank(P)$ or $rank(M^{j}_{ia}) \le
rank(I-P)$, which implies $ker(M^{j\dagger}_{ia}) \neq \emptyset$.
So at stage (2.3) the state $|\xi^{j}_{ia}\rangle$ can always be
found.

Consider an arbitrary iteration of stage 2 with $i<m$. First, it
is necessary to prove that $M^{j}_{ia}$, $\lambda^{j}_{ia}$, and
$|\phi^{j}_{ia}\rangle$ satisfy the condition
\begin{equation}
M^{j\dagger}_{ia}M^{j}_{ia}-\lambda^{j}_{ia}|\phi^{j}_{ia}\rangle\langle\phi^{j}_{ia}|
\ge 0 \label{eq:cond}
\end{equation}
so that the function $f$ can be applied to them at stage (2.2).
This will be proved later. Now we assume that it holds and then by
Eq.(\ref{eq:lem}) get
\begin{equation}
M^{j\dagger}_{ia}|\theta^{j}_{ia}\rangle=\sqrt{\mu^{j}_{ia}}|\phi^{j}_{ia}\rangle.
\label{eq:mt}
\end{equation}
By Eq.(\ref{eq:psijia}), Eq.(\ref{eq:mt}) and
$M^{j\dagger}_{ia}|\xi^{j}_{ia}\rangle=0$ we have
\begin{equation}
\begin{array}{l}
M^{j\dagger}_{ia}|\psi^{j}_{ia}\rangle\langle\psi^{j}_{ia}|M^{j}_{ia}=\lambda^{j}_{ia}|\phi^{j}_{ia}\rangle\langle\phi^{j}_{ia}|.
\end{array}
\label{eq:mjia}
\end{equation}
Then it follows that the probability of obtaining the measurement
outcome corresponding to $|\psi^{j}_{ia}\rangle$ is
\begin{equation}
\begin{array}{l}
|\langle\psi^{j}_{ia}|\psi'\rangle|^{2}
=\langle \psi |M^{j\dagger}_{ia}|\psi^{j}_{ia}\rangle\langle\psi^{j}_{ia}|M^{j}_{ia}|\psi\rangle/\|M^{j}_{ia}|\psi\rangle\|^{2}\\
=\lambda^{j}_{ia}\langle\psi|\phi^{j}_{ia}\rangle\langle\phi^{j}_{ia}|\psi\rangle
/\|M^{j}_{ia}|\psi\rangle\|^{2}.
\end{array}
\end{equation}
Taking into account the prior probability of reaching this stage
$\|M^{j}_{ia}|\psi\rangle\|^{2}$, the probability of the protocol
stopping at stage (2.4) with the current values of $i$, $j$ and
$a$ is
$\lambda^{j}_{ia}\langle\psi|\phi^{j}_{ia}\rangle\langle\phi^{j}_{ia}|\psi\rangle$.

Therefore, the total probability of the protocol yielding the
final outcome $i$ is
\begin{equation}
\sum\limits_{a=0}^{1}\sum\limits_{j=1}^{d_{ia}}{\lambda^{j}_{ia}\langle\psi|\phi^{j}_{ia}\rangle\langle\phi^{j}_{ia}|\psi\rangle}
=\sum\limits_{a=0}^{1}{\langle \psi |E_{ia}| \psi \rangle}
=\langle \psi | E_{i} | \psi \rangle
\end{equation}
for all $i<m$. And naturally the probability of yielding the final
outcome $m$ will be $1-\sum_{i=1}^{m-1}{\langle \psi | E_{i} |
\psi \rangle}=\langle \psi | E_{m} | \psi \rangle$. So this
protocol realizes the POVM $\{E_1,\dots,E_m\}$.

Now we go back to prove that the condition (\ref{eq:cond}) is
always fulfilled when $i<m$. Actually, if
\begin{equation}
M^{j\dagger}_{ia}M^{j}_{ia}=\sum_{j'=j}^{d_{ia}}{\lambda^{j'}_{ia}|\phi^{j'}_{ia}\rangle\langle\phi^{j'}_{ia}|}
+\sum_{i'=i+1}^{m}\sum_{j'=1}^{d_{i'a}}{\lambda^{j'}_{i'a}|\phi^{j'}_{i'a}\rangle\langle\phi^{j'}_{i'a}|}
\label{eq:mm}
\end{equation}
holds, then the inequality (\ref{eq:cond}) will be true.

We will prove Eq.(\ref{eq:mm}) by induction on the indices
$(i,j)$. We consider only the case of $a=0$, because the case of
$a=1$ can be dealt similarly. At the beginning, $(i, j)=(1, 1)$,
$M^{1}_{10}=P$. It follows from Eq.(\ref{eq:spec}) and
$\sum_{i=1}^{m}E_{i0}=P$ that
\begin{equation}
\sum_{i=1}^{m}\sum_{j=1}^{d_{i0}}{\lambda^{j}_{i0}|\phi^{j}_{i0}\rangle\langle\phi^{j}_{i0}|}=P.
\end{equation}
So Eq.(\ref{eq:mm}) holds. Now suppose that for some indices $(i,
j)$, Eq.(\ref{eq:mm}) is valid. If $j<d_{i0}$, then by
Eq.(\ref{eq:2.5.1}) and Eq.(\ref{eq:mjia}) we have
\begin{equation}
\begin{array}{l}
M^{(j+1)\dagger}_{i 0}M^{j+1}_{i 0} =M^{j\dagger}_{i 0}M^{j}_{i 0}
-M^{j\dagger}_{i 0}|\psi^{j}_{i 0}\rangle\langle\psi^{j}_{i 0}|M^{j}_{i 0}\\
=\sum\limits_{j'=j+1}^{d_{i 0}}{\lambda^{j'}_{i 0}|\phi^{j'}_{i
0}\rangle\langle\phi^{j'}_{i 0}|}
+\sum\limits_{i'=i+1}^{m}\sum\limits_{j'=1}^{d_{i'0}}{\lambda^{j'}_{i'0}|\phi^{j'}_{i'0}\rangle\langle\phi^{j'}_{i'0}|},
\end{array}
\end{equation}
which implies that Eq.(\ref{eq:mm}) is also valid for the next
indices $(i, j+1)$. Similarly, if $j=d_{i0}$, then by
Eq.(\ref{eq:2.5.2}) and Eq.(\ref{eq:mjia}) the validity of
Eq.(\ref{eq:mm}) for the next indices $(i+1, 1)$ can be proved.

To analyze the efficiency of our protocol, we should be aware that
its basic idea is to individually realize each item
$\lambda^{j}_{ia}|\phi^{j}_{ia}\rangle\langle \phi^{j}_{ia}|$ in
Eq.(\ref{eq:spec}) and contribute it to the corresponding POVM
element $E_{i}$ for all $i<m$, while leaving the residual
probability to $E_m$. Since each item needs exactly a projective
measurement, our protocol performs at most $\max\limits_{a=0,1}
\{\sum\limits_{i=1}^{m-1}{rank(E_{ia})}+1\}$ projective
measurements in total. Actually we can rearrange the POVM elements
$\{E_1,\dots,E_m\}$ to minimize this upper bound.

Summarizing, we get the following theorem:
\begin{theorem}
A POVM $\{E_1,...,E_m\}$ can be realized by a sequence of
projective measurements on the original space if and only if there
exists a projection operator $P \neq I$ such that $[E_i,P]=0$ for
$i=1,\dots,m$. \label{thm}
\end{theorem}

As an application, we consider the problem of unambiguous
discrimination \cite{IDP} of mixed quantum states. Suppose a state
is secretly chosen from two quantum states $\rho_1$ and $\rho_2$
whose supports have nonempty intersection, i.e. $supp(\rho_1) \cap
supp(\rho_2) \neq \emptyset$. Choose a state $|\psi\rangle \in
supp(\rho_1) \cap supp(\rho_2)$. If a POVM $\{E_0,E_1,E_2\}$ can
be used to unambiguously distinguish the two states (where $E_1$,
$E_2$ correspond to $\rho_1$, $\rho_2$ respectively, and $E_0$
leads to no conclusion), the condition
\begin{equation}
tr(E_1\rho_2)=tr(E_2\rho_1)=0
\end{equation}
should be fulfilled. Then we have $supp(\rho_2) \subset ker(E_1)$
and $supp(\rho_1) \subset ker(E_2)$, which implies $|\psi\rangle
\in ker(E_1)$ and $|\psi\rangle \in ker(E_2)$. Let
$P=|\psi\rangle\langle\psi|$. Then one can verify that $[E_i,P]=0$
for $i=0,1,2$. Hence this POVM can be realized by our approach.

A surprising consequence of theorem \ref{thm} is that when
allowing sequences of projective measurements, an arbitrary POVM
can be realized by introducing only a single ancillary dimension,
as the following corollary states:
\begin{corollary}
An arbitrary POVM on a $d$-dimensional space can be realized by a
sequence of projective measurements on an extended
$(d+1)$-dimensional space.
\end{corollary}
To prove this, note that a POVM $\{E_1,\dots,E_m\}$ on a
$d$-dimensional Hilbert space ${\cal H}$ can be mapped to the POVM
$\{E_1,\dots,E_m,|\psi\rangle\langle\psi|\}$ on any
$(d+1)$-dimensional space ${\cal H'}$ formed by adding an extra
basis element $|\psi\rangle$ to ${\cal H}$. Letting
$P=|\psi\rangle\langle\psi|$, one can find that our condition
holds trivially. So we can realize this POVM by utilizing our
protocol on the extended space ${\cal H'}$. Note that in this
situation, one never needs to perform the projective measurement
$\{P,I-P\}$ at stage 1, since only the outcome corresponding to
$I-P$ can be obtained. We can directly set $a=1$. The other part
of the protocol remains the same.

In conclusion, we present a protocol that can realize a class of
POVMs by performing a series of projective measurements on the
original system, in the sense that it can simulate the probability
distribution of the measurement outcomes for any input state. A
necessary and sufficient condition for a POVM to be realizable in
this way is also derived. Our method requires no auxiliary system
and thus may be easier to implement in practice than the one
provided by Neumark's theorem. Moreover, arbitrary POVMs can be
realized by adopting our protocol on an extended space which is
formed by introducing only a single extra dimension. Our work may
help with the implementation of generalized quantum measurements
in the tasks where only the measurement outcome is concerned such
as quantum state estimation and discrimination.

We gratefully thank Michael Hall for suggesting the elegant
expression of our theorem using the Lie bracket [,] and also
pointing out corollary 1. This work was supported by the Natural
Science Foundation of China (Grants Nos. 60503001, 60321002 and
60305005).

\emph{Appendix}: Here we prove lemma {\ref{lemma1}}. Suppose $M$
has singular value decomposition
$M=\sum_{i=1}^{d}{p_i|\psi_i\rangle\langle\phi_i|}$, where
$p_i>0$, $\langle \psi_i|\psi_j \rangle=\delta_{ij}$, $\langle
\phi_i|\phi_j \rangle=\delta_{ij}$, and $d=rank(M)$. Then we have
$M^{\dagger}M$ has spectral decomposition
$M^{\dagger}M=\sum_{i=1}^{d}{p^{2}_i|\phi_i\rangle\langle\phi_i|}$.
By $M^{\dagger}M-\lambda|\phi\rangle\langle\phi| \ge 0$, we
suppose $M^{\dagger}M$ can also be written in the form
$M^{\dagger}M=\lambda|\phi\rangle\langle\phi|+\sum_{i=1}^{m}{q_i|\zeta_i\rangle\langle\zeta_i|}$
for some $m \ge d-1$, $q_i>0$ and $|\zeta_j\rangle$. Then by
theorem 2.6 of Ref.\cite{NC00}, we conclude that there exists a
$(m+1) \times (m+1)$ unitary matrix $U=(u_{ij})_{i,j=1,...,m+1}$
such that
$\sqrt{\lambda}|\phi\rangle=\sum_{i=1}^{d}{u_{1i}p_i|\phi_i\rangle}$,
where $\|u\|^2=\sum_{i=1}^{d}{|u_{1i}|^{2}} \le 1$. Let
$|\theta\rangle=\frac{1}{\|u\|}\sum_{i=1}^{d}{u_{1i}|\psi_i\rangle}$,
$\mu=\lambda/\|u\|^2 \ge \lambda$. Then we obtain
$M^{\dagger}|\theta\rangle=\frac{1}{\|u\|}\sum\limits_{i=1}^{d}{u_{1i}p_i|\phi_i\rangle}=\sqrt{\mu}|\phi\rangle$.
$\square$

\end{document}